\documentclass[pdflatex,sn-mathphys-num]{sn-jnl}

\usepackage{graphicx}%
\usepackage{multirow}%
\usepackage{amsmath,amssymb,amsfonts}%
\usepackage{amsthm}%
\usepackage{mathrsfs}%
\usepackage[title]{appendix}%
\usepackage{xcolor}%
\usepackage{textcomp}%
\usepackage{manyfoot}%
\usepackage{booktabs}%
\usepackage{algorithm}%
\usepackage{algorithmicx}%
\usepackage{algpseudocode}%
\usepackage{listings}%
\usepackage{subcaption}%

\theoremstyle{thmstyleone}%
\theoremstyle{thmstyletwo}%

\theoremstyle{thmstylethree}%

\begin{document}

\title[Article Title]{A hard energy Inverse Compton X-ray source driven by a laser-plasma accelerator used to inspect additively manufactured components.}




\author[1,*] {C. Thornton}

\author[2] {S. Karimi}

\author[2] {S. Glenn}
\author[2] {W. D. Brown}
\author[2] {N. Dragani\'{c}} 
\author[2]{ M. Skeate} 
\author[2] {M. Ferrucci} 

\author[3] {Q. Chen}
\author[3] {R. Jacob}
\author[3] {K. Nakamura}
\author[3] {T. Ostermayr}
\author[3]  {J. van Tilborg}

\author[1]{C. Armstrong}
\author[1]{O. J. Finlay}

\author[4] {N. Turner}
\author[5] {S. Glanvill}

\author[2] {H. Martz}
\author[3] {C. Geddes}

\affil[1]{STFC Rutherford Appleton Laboratory,  Central Laser Facility, Didcot, UK}
\affil[2]{Lawrence Livermore National Laboratory, Nondestructive Characterization Institute,  Livermore, USA}
\affil[3]{Lawrence Berkeley National Laboratory, Accelerator Technology and Applied Physics, Berkeley, USA}
\affil[4]{The Manufacturing Technology Centre, Verification \& Validation, Coventry, UK}
\affil[5]{Renishaw plc, Materials Analysis and Research, Wotton-under-Edge, UK}


\abstract{For the first time, we demonstrate the application of an inverse Compton scattering X-ray Source, driven by a laser-plasma accelerator, to image an additively manufactured component. X-rays with a mean energy of 380 keV were produced and used to image an additively manufactured part made of an Inconel (Nickel 718) alloy. Because inverse Compton scattering driven by laser-plasma acceleration produces high-energy X-rays while maintaining a focal spot size on the order of a micron, the source can provide several benefits over conventional X-ray production methods, particularly when imaging superalloy parts, with the potential to revolutionise what can be inspected.}

\keywords{X-ray Generation, Inverse Compton Scattering, Tomography, Additive Manufacturing, Laser Plasma Acceleration, Inconel (Nickel 718)}



\maketitle

\section*{Introduction}

In additive manufacturing (AM) technologies, such as 3D printing, components are built up layer-by-layer, unlike traditional subtractive manufacturing, where material is removed from a block to form a component. AM parts are often high-value, complex in geometry, and can be made from superalloys. They can be used in a variety of challenging environments such as engine manifolds in the aerospace industry \cite{BLAKEYMILNER2021}, a range of lighter and more complex components for the automotive industry \cite{ZHAO2023}, and potentially as plasma divertors in fusion reactors \cite{JeongHa2023}. 
\\
\\
The advent of AM has led to two key capabilities in manufacturing. Firstly, there is a more widespread use of superalloys, such as Inconel (Nickel 718), which are more suited to this type of manufacturing and can withstand extremes of temperature and pressure \cite{Nordin_2017}. Secondly, AM allows designers to optimise components to have maximal structural integrity whilst using the minimum amount of material, or to add increased functionality (such as internal cooling channels) \cite{BLAKEYMILNER2021}. As a result of these two capabilities, components have greater complexity and are made with a material of high atomic number, which makes inspection difficult. Inspecting AM components, such as automotive heat exchangers, is important as even minuscule defects ($<10 \mu m$) could lead to part failure \cite{Niknam2020}. The bottleneck to the wide adoption of this manufacturing technique is validating component integrity of high atomic metal parts via nondestructive characterization (NDC), the preferred approach being tomographic imaging with X-rays \cite{Withers2021}.
\\
\\
X-ray computed tomography (XCT) involves taking several X-ray images of a part from different viewing angles, from which a 3D image can be reconstructed and interrogated for flaws. Current methods for producing X-rays in industry are hemmed in by two technologies: the X-ray tube, and the linear accelerator (Linac) \cite{Martz2016}. X-ray tubes with tens of micrometre focal spot size can be used to acquire sub-millimetre spatial resolution images. However, energies are under 600 kV, limiting the atomic number and thickness of materials that can be inspected. Linacs produce X-rays with energies in the MV range but have 0.7-2 mm focal spots, and hence yield insufficient spatial resolution for structure metrology and defect detection \cite{Thompson_2016}. These conventional X-ray sources are thus limited in spatial resolution for applications demanding higher photon energies, such as imaging high atomic number metal parts. In both conventional X-ray production methods, electrons are accelerated and then collide with a metal target to generate broad-spectrum bremsstrahlung X-rays (in addition to characteristic X-ray peaks that are dependent on the target material). The energy referenced in these sources is the maximum energy, while almost all photons are at substantially lower energy in a distribution with 100\% spread peaked in intensity at the lower energies. As a result, these bremsstrahlung sources are polychromatic (have a wide energy spectrum), which degrades image quality when scanned objects harden the X-ray beam and limit our ability to infer physical information from the reconstructed image \cite{Azevedo2016,Martz2017,Withers2021}. Finally, X-ray beam energy tunability is a desirable property to optimize the imaging quality for objects made of different materials. The simultaneous need for sub-mm spatial resolution, MV X-ray energies, narrow spectra and tunability necessitates the use of another X-ray source technology, for instance, an inverse Compton scattering (ICS) X-ray Source. 
\\
\\
In an X-ray ICS, low-energy photons from a laser pulse interact with a high-energy electron beam to generate X-rays. This method can generate MV X-rays from an electron beam with energies on the order of 100 MeV. However, conventional accelerators required to generate such high-energy electron beams are large \cite{Weller2009}, limiting their use. An alternative approach to electron beam production is to use a laser-plasma accelerator (LPA) \cite{Mangles2004,Geddes2004,Faure2004}. An LPA uses a short pulse (45 fs), high energy (1-2 J) laser which is focused to a high intensity ( $10^{18}-10^{19} W/cm^2$) inside a gas, generating a plasma. The laser-plasma interaction sets up a trailing plasma wakefield behind the laser pulse, within which electron bunches become trapped. Electrons in this wakefield are accelerated to relativistic energies within distances of 1-2 centimetres. This allows the accelerator to be a thousand orders of magnitude smaller than conventional accelerators. The electron beam created is typically a few micrometres\cite{Bingham2014}, and can be as small as 0.1 micrometre \cite{Plateau2012}. 
\\
\\
Generating X-rays using LPA-ICS \cite{HaiEn2015, Ta_Phuoc2012, GEDDES2015116} would offer major advantages over the current MV X-ray production techniques for XCT. Firstly, with the LPA-ICS X-ray production technique, the X-ray energies are tunable between 100 keV and 1 MeV (or potentially greater if needed), making the widespread industrial inspection of complex components built using superalloys possible. More importantly, the LPA-ICS X-ray source has a small focal spot size, between 0.1 and 10 $\mu m$ \cite{Kramer2018}, which would theoretically allow micrometre spatial resolution \cite{Rykovanov_2014} to be achieved (cf. the mm resolution achieved at MeV X-ray energies produced by linear accelerators) across these X-ray energies. The combination of these two key factors, small focal spot and the generation of MeV X-rays, offers the prospect of achieving sub-micrometre resolution X-ray images of high-density alloy AM components, which can be used to generate high spatial resolution XCT images. X-rays using LPA-ICS have been used for radiography \cite{Döpp_2016,Haden2016,CHEN2016} and in keV X-ray tomography \cite{Ma2020} but have never been used to meet the new challenges presented by modern AM manufacturing techniques to the NDC community. 
\\
\\
This paper reports on the first demonstration of using a high-energy X-ray beam generated by an LPA-ICS technique to achieve an XCT of a superalloy AM part made from Nickel 718 (Inconel). X-ray images were acquired and used to produce a three-dimensional tomographic reconstruction of the part. Analyses of the reconstructed image allowed us to confirm the design parameters of the AM part, showing that the basic building blocks of this X-ray imaging technique are present and can be refined to meet this new application.


\section*{Method}
\label{sec:Method}

\subsection*{Beam Line Set-up and Data Acquisition}
\label{subsec:BeamLine}

\begin{figure}[h]
\centering
\includegraphics[width=\linewidth]{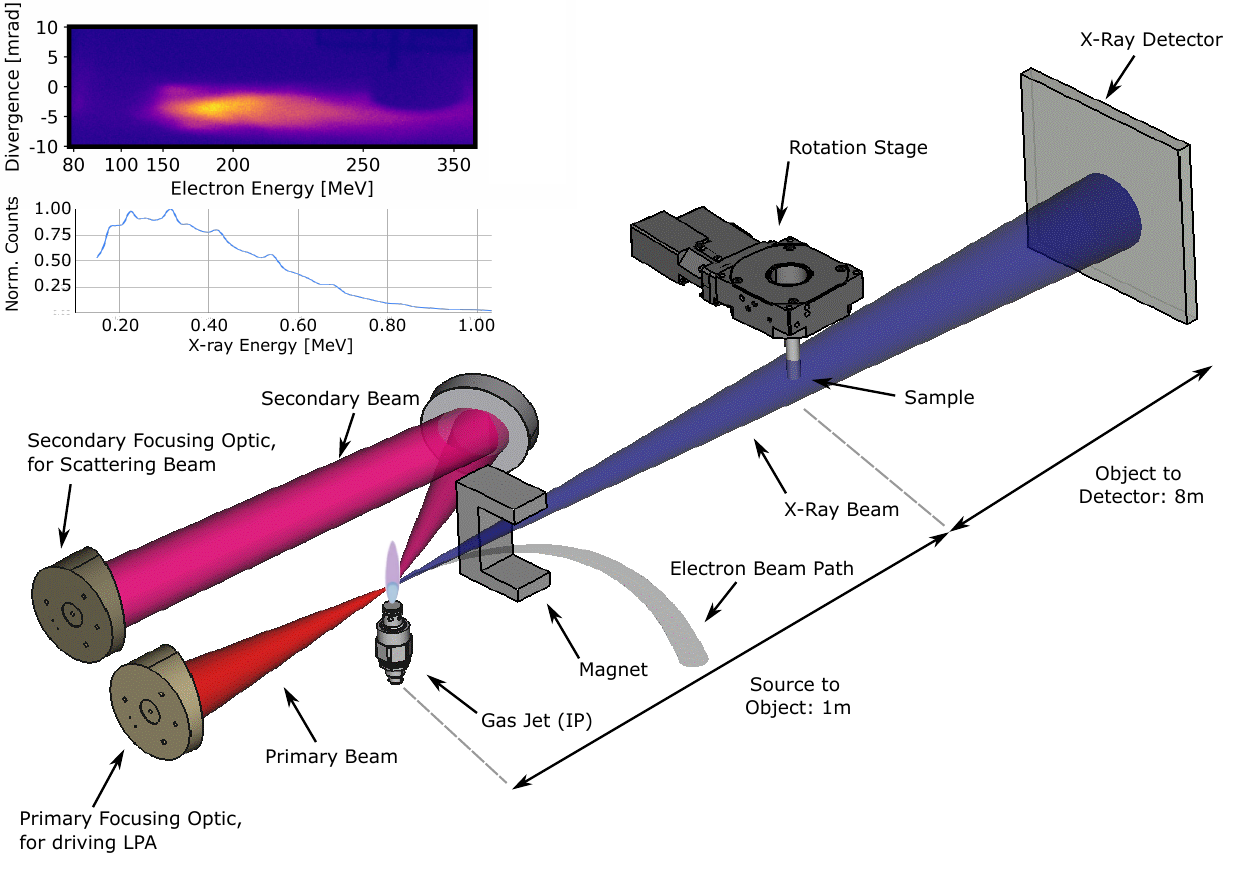}
\caption{Schematic of the BELLA HTW set-up, showing the primary and secondary scattering laser beam, along with the positions of the gas jet, sample, and X-ray detector. The schematic is an illustration only and does not represent the real dimensions of the set-up. The graphs in the top left corner are a typical electron and calculated ICS X-ray spectrum.}
\label{fig:Layout}
\end{figure}

\begin{figure}[t]
     \centering
     \begin{subfigure}[b]{0.45\textwidth}
         \centering
         \includegraphics[width=2in, height = 2in]{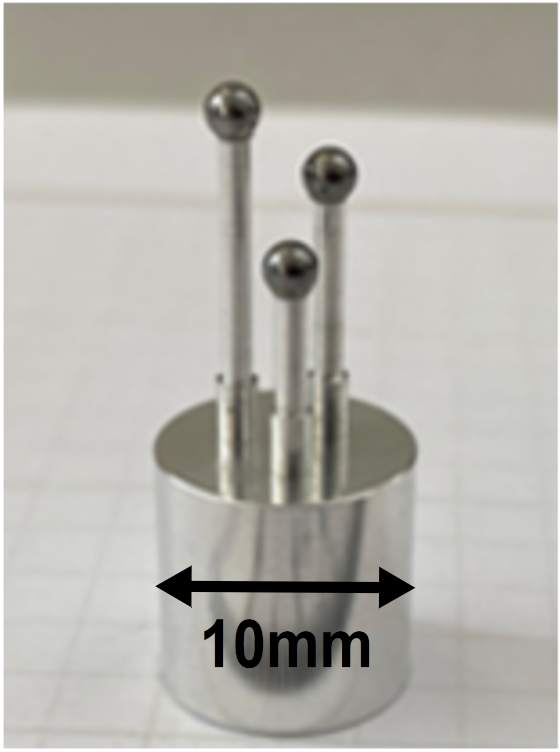} 
         \caption{}
         \label{fig:Phantom}
     \end{subfigure}
     \hfill
     \begin{subfigure}[b]{0.45\textwidth}
         \centering
         \includegraphics[width=2in, height = 2in]{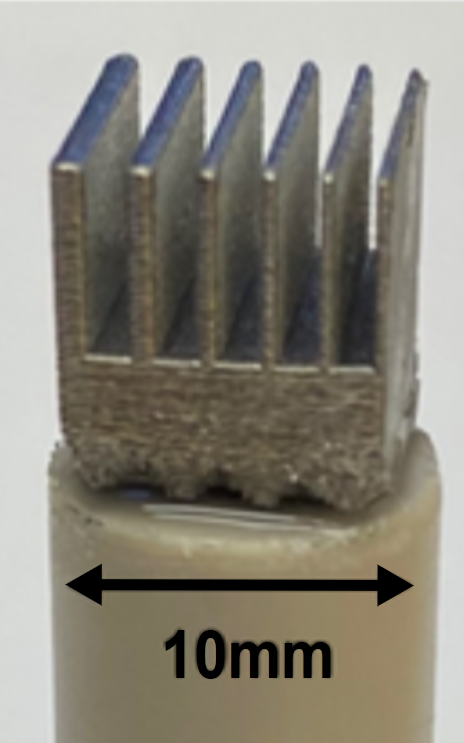}
         \caption{}
         \label{fig:AMpart}
     \end{subfigure}
     \caption{The two samples that were scanned using the ICS source. Figure \ref{fig:Phantom} is the calibration phantom comprising of aluminium base with 2.5 mm O.D. tungsten carbide (WC) spheres. Figure \ref{fig:AMpart} is the AM printed Inconel part with varying fin thickness.}
        \label{fig:part}
\end{figure}

The experiment was carried out at the Lawrence Berkeley National Laboratory (LBNL), Berkeley Lab Laser Accelerator (BELLA) Hundred Terawatt Laser Facility (HTW), which is a Ti:Sapphire chirped pulse amplification (CPA) laser providing two beams into the experimental area \cite{Ostermayr2020,Chen2023}. The dual-beam design of this laser system gives flexibility to ICS setup, allowing for the control of the X-ray beam properties, such as the width of the X-ray spectrum. For this experiment, the primary beam delivered a $\approx$ 600 mJ, 40 fs laser pulse, while the secondary beam provided a 300 mJ, 40 fs laser pulse. Both beams were delivered at a repetition rate of 1 Hz into the same target chamber and were independently compressed to their final pulse lengths by separate compressors. Both lasers are focused by independent off-axis paraboloid mirrors. The primary beam is focused at 1.5 m focal length, to give an 18 $\mu$m focal spot and directed towards a magnetic spectrometer. The secondary beam is focused in a counter-propagating direction to the main beam, at a focal length of 1.1 m, to give a matched focal spot size of 18 $\mu$m. The secondary beam crosses the main beam at an angle of approximately 16\textdegree. The laser system is equipped with sophisticated controls and diagnostics that allow for the accurate placement and monitoring of the two beams. The general layout of the system can be seen in Figure \ref{fig:Layout}.
\\
\\
The primary beam was focused on a gas jet \cite{Ocean2021} placed in the centre of the target chamber, which is maintained at a vacuum of $10^{-5}$ torr. The intensity of the primary beam is high enough to generate a plasma wakefield as described above. Electrons were generated in a stable regime with a peak energy of 125 MeV, a charge of 20 pC was generated and an energy spread of around 50 \%. The secondary beam was used as the scattering pulse, to generate X-rays via ICS from the electrons with a mean X-ray energy of 380 keV. After the interaction point (IP), the remaining primary laser beam, electrons, and X-rays co-propagate towards the sample, the primary laser beam is removed using a thin aluminium foil (not shown in Figure \ref{fig:Layout}) and the electrons are deflected by a fixed dipole field magnet approximately 40 cm from the IP. The deflected electrons are sent to a phosphor screen and the magnitude of their deflection is used to calculate the electron beam energy. The X-rays were then propagated through the sample (as shown in Figure \ref{fig:Layout}) which is placed approximately 1 m from the IP. The sample was placed on a Newport vacuum-compatible rotation stage with 360° continuous motion. For ease of data collection, the rotation stage was mounted on two 75 mm travel linear Zaber stepper stages, which allowed the sample to be moved in and out of the X-ray beam. Furthermore, two Newport high precision, miniature, piezo sample stages were placed on top of the rotation stage which allowed the centre of rotation of the sample to be easily found. The samples were glued onto a polymer post that could then be held by a custom clamp, which was attached to the top of the two linear stages. The X-rays exited through a thin aluminium window in the target chamber and then propagated a further 8 m to a Varex Imaging 1611 AP26 amorphous silicon flat panel detector. The 4096 × 4096 pixel detector has a gadolinium oxysulphide (GOS) scintillator and 0.1 mm × 0.1 mm pixel size. Considering geometric magnification, the effective detector pixel size at the centre of rotation is 0.011 mm × 0.011 mm.
\\
\\
Two samples (Figure \ref{fig:part}) were scanned in this system, a calibration phantom comprising tungsten carbide spheres supported by aluminium stems fabricated at Lawrence Livermore National Laboratory and an AM Inconel part fabricated by Renishaw. A tomographic scan of a sample is acquired by placing it on a rotating stage and taking X-ray images at different angular positions called projections. To increase the collected flux, and consequently improve the signal-to-noise ratio (SNR), multiple X-ray images were taken and averaged at each projection. In each projection, pixels were binned 4 x 4 to increase the signal-to-noise ratio and to reduce read-out time overhead. Consequently, the effective detector pixel size at the centre of rotation was 0.05 mm × 0.05 mm.
\\
In addition to the object projections, we acquire offset and gain calibration data. Offset (interchangeably called dark current) data characterises the output of each detector pixel without X-ray exposure. Gain (also called open-field) data measures the output of each pixel with X-rays but no object. The offset calibration data is used to correct the pixel-wise variable dark current of the object X-ray data. The gain calibration data is used to correct variable gains and in the application of the Beer-Lambert law. Two frames were acquired per second, one of which had an X-ray pulse and the other was offset data. Offset frames were collected frequently because the dark current drifted during the scan (shown in Figure \ref{fig:drift}). A gain calibration was taken before the object scan: the data set comprised of multiple X-ray images with interleaved offset frames.
\\
\\
The calibration phantom was used to test our acquisition technique and confirm the geometry of the imaging system. As a result of it being symmetrical, only 60 frames at 25 projections were taken over an angular range of 288 \textdegree. The AM part was much more intricate and non-symmetrical, as a result, we took 40 frames at 301 projections over 360 \textdegree.  The reduction in the number of X-ray images per projection of the calibration phantom allowed for the data to be collected in a few hours, making repeat data sets possible.

\subsection*{Data Processing and Image Reconstruction}
\label{subsec:DataProcessing}

\begin{figure}[h]
     \centering
     \begin{subfigure}[b]{0.49\textwidth}
         \centering
         \includegraphics[width=2in,height = 2in]{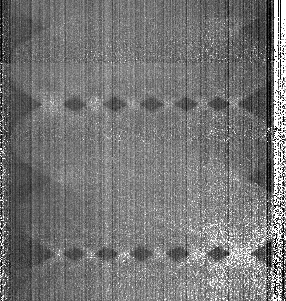} 
         \caption{}
         \label{fig:sino}
     \end{subfigure}
     \hfill
     \begin{subfigure}[b]{0.49\textwidth}
         \centering
         \includegraphics[width=2in, height = 2in]{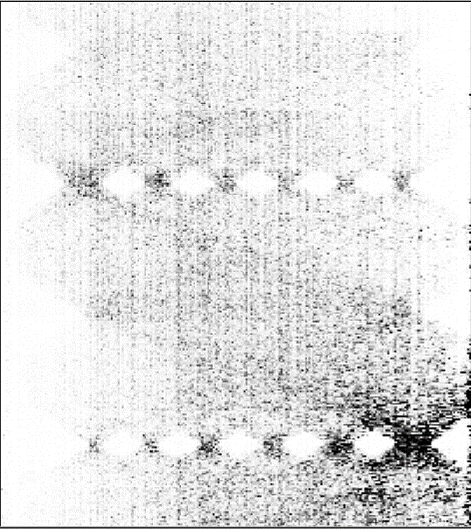} 
         \caption{}
         \label{fig:weights}
     \end{subfigure}
\caption{  Figure \ref{fig:sino}  shows a sinogram of the AM part. Weights in the RWLS reconstruction algorithm for the reconstruction AM part are shown in Figure \ref{fig:weights}. Brighter values represent higher weights. The degraded regions of the sinogram are de-emphasised by the black areas in Figure \ref{fig:weights}.}
\label{fig:sinogram}
\end{figure}

Consider a single radiography called an object frame, represented by $I (\upsilon,j,m)$ where $\upsilon$ represents the view angle, $j$ represents the frame at that view and $m$ represents the pixel. We subtract the offset data $I_d$ and average the frames to generate object projections as follows:

\begin{equation}\label{eqn2}
   \overline{I}(\upsilon,m) =  \frac{1}{F_p}  \sum_{j=0}^{F_p}  I (\upsilon,j,m) - I_d (\upsilon,j,m),
\end{equation} 
\\
\\
where ${F_p}$ represents the number of frames at each view. Similarly, we subtract the offset data $I_d$ and average the gain calibration frames to produce a gain calibration table.  

\begin{equation}\label{eqn1}
   \overline{I}_{0} (m) =  \frac{1}{F_\alpha}  \sum_{j=0}^{F_\alpha}  I_{0} (j,m) - I_d (j,m) ,
\end{equation} 
\\
\\
where $I_0$ represents the gain calibration data and ${F_\alpha}$ represents the number of frames in the gain scan. The Beer-Lambert law \cite{BeerLambert} is then applied to give: 





\begin{equation}\label{eqn:lambert}
   p(\upsilon,m) = \ln \frac{\overline{I}_{0}(m)}{\overline{I}(\upsilon)(m)} 
\end{equation} 
\\
\\
The contrast-to-noise ratio varied throughout the scans, and there were noise spikes in the data. We, therefore, chose to use a statistical image reconstruction method which handles these noise and variability problems better than filtered back projection. The statistical reconstruction algorithm solves a regularized weighted least squares problem shown in equation \ref{eqn:lsq}, where we solve for image parameters $\mathop{x{}}\limits^\wedge$. The system matrix $A$ represents the forward projection, $P$ represents the set of all measured projections, $p(\upsilon,m)$. The weights matrix $W$ suppresses unreliable data and the strength of regularization is controlled by a scalar, $\beta$. We use the total variation norm (TV) for regularization.  


\begin{equation}\label{eqn:lsq}
   \mathop{x{}}\limits^\wedge  =  \min_{x} (Ax - P)^T W(Ax - P) + \beta \| x \|^{2}_{T V}
\end{equation} 
\\
\\
We designed weighting for the statistical reconstruction according to the quality of the data in each projection. A sinogram of one slice of the AM part is shown in Figure \ref{fig:sino}. We see noise increase as attenuation increases, as expected in X-ray imaging, but we also see severe degradation in the lower right region of the sinogram. Figure \ref{fig:weights} shows the weights used in the reconstruction of the AM part. These weights are a sigmoid function of log attenuation, followed by grayscale erosion and are calculated using:

\begin{equation}\label{eqn6}
   W(\rho) = 1 -  \frac{1}{1+e^{c\rho+d}},
\end{equation} 
\\
\\
where $\rho = p(\upsilon,m)$ represents log-attenuation for a single projection ray. The constants $c = -5$ and $d =5.2$ were empirically determined. The weights for the reconstruction of the calibration phantom de-emphasized later projections because the quality of the projections in that scan degraded over time (not shown here). The final reconstruction and rendering of the samples were performed using the Livermore Tomography Tools (LTT) software\cite{Kyle2022}. 
\\
\\


\section*{Results}
\label{sec:Results}

\begin{figure}[h]
     \centering
     \begin{subfigure}[b]{0.3\textwidth}
         \centering
         \includegraphics[width=1.5in, height = 1.5in]{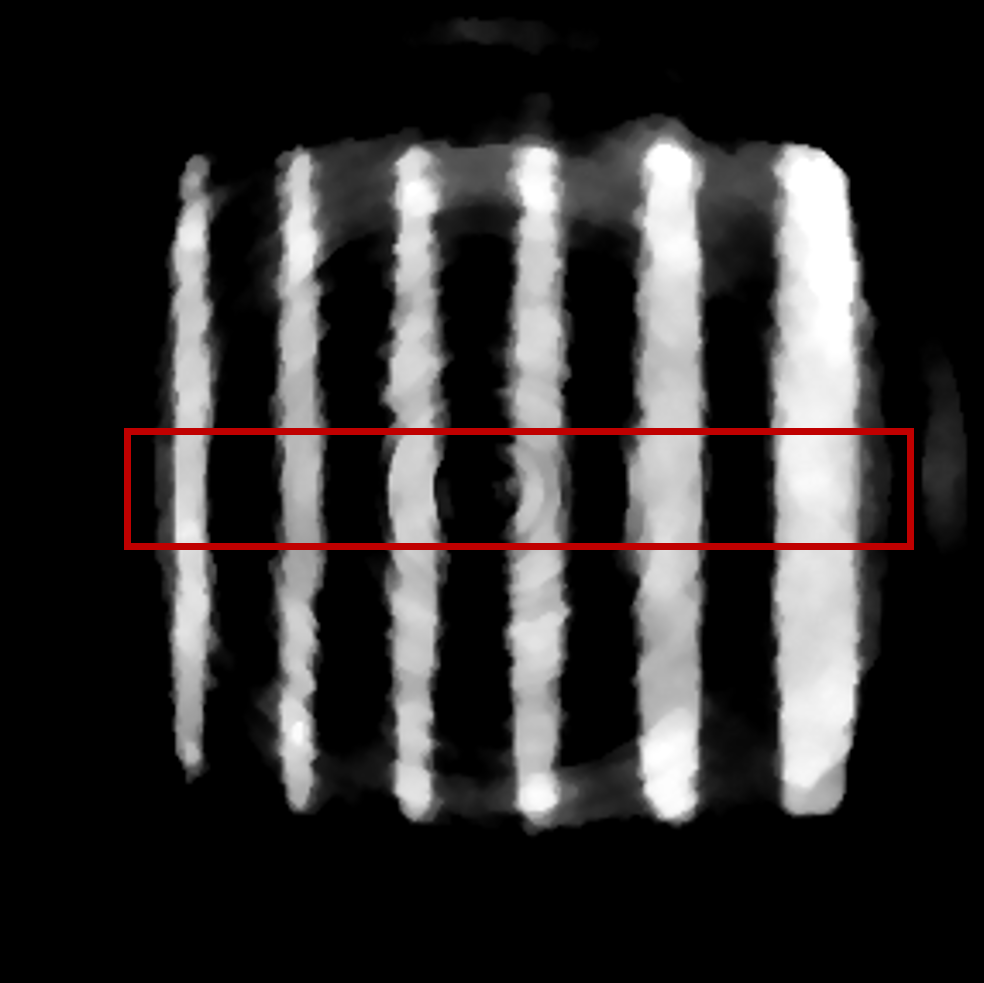}
         \caption{}
         \label{fig:CTRenSlice}
     \end{subfigure}
     \hfill
     \begin{subfigure}[b]{0.38\textwidth}
         \centering
         \includegraphics[width=1.5in, height = 1.5in]{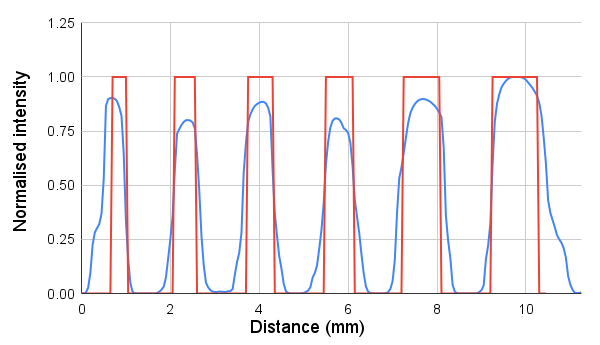}
         \caption{}
         \label{fig:RenLineout}
     \end{subfigure}
     \hfill
     \begin{subfigure}[b]{0.3\textwidth}
         \centering
         \includegraphics[width=1.5in, height = 1.5in]{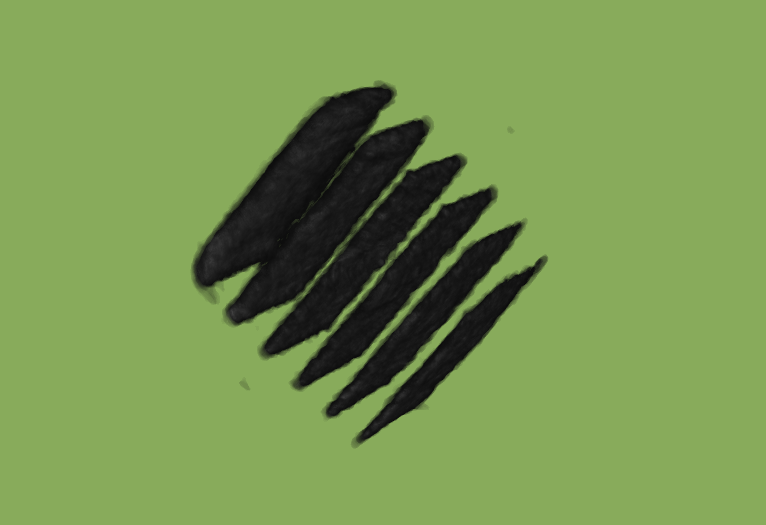} 
         \caption{}
         \label{fig:RenReconstruct}
     \end{subfigure}
     \caption{Figure \ref{fig:CTRenSlice} shows a XCT slice through the reconstruction image of the AM part. Figure \ref{fig:RenLineout} shows the average line out, with the blue line representing the XCT data and the red line showing the destructive measuring technique. Figure \ref{fig:RenReconstruct} shows a 3D-rendered image of the AM part, with the fins clearly visible.}
        \label{fig:Renishaw}
\end{figure}

This results section will focus on the X-ray imaging results of the AM object, which is the focus of this paper. A brief description of the X-ray properties obtained has been reported here for completeness but as a standard ICS set-up was used, much greater detail on the X-ray source can be found in other papers \cite{Rykovanov_2014}.  
\\
\\
A typical electron spectrum is shown in Figure \ref{fig:Layout}, along with ICS X-ray spectra that were estimated by the method described by Rykovanov\cite{Rykovanov_2014}. This calculation yields on-axis photon energies using measured electron energies while accounting for experimental conditions. The mean energy of this estimated X-ray spectrum is 380 keV. Although the spectrum is very broad it is significantly different to a Bremsstrahlung spectrum, in the sense that the peak of the spectrum is close to the mean energy value. A Bremsstrahlung source is peaked at the lower energy of the spectrum. This is a key difference in the sources and means the ICS X-ray source is made up of more high-energy X-rays that contribute to the image rather than low energy X-rays which cause beam hardening, the consequences of which will be discussed later in the paper. 
\\
\begin{figure}[t]
     \centering
     \begin{subfigure}[b]{0.3\linewidth}
         \centering
         \includegraphics[width=1.5in, height = 1.5in]{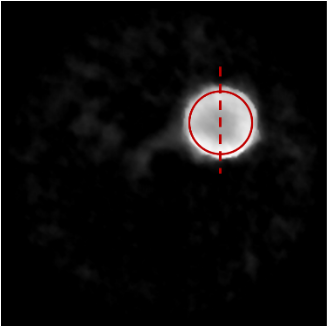}  
         \caption{}
         \label{fig:BallBearing}
     \end{subfigure}
     \begin{subfigure}[b]{0.3\linewidth}
         \centering
         \includegraphics[width=1.5in, height = 1.5in]{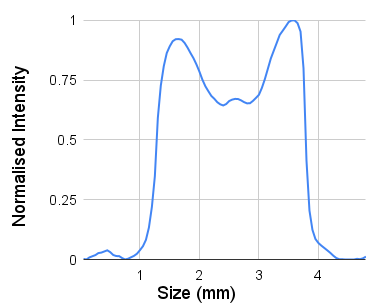}
         \caption{}
         \label{fig:lineout}
     \end{subfigure}
     \begin{subfigure}[b]{0.38\linewidth}
         \centering
         \includegraphics[width=2in, height = 1.5in]{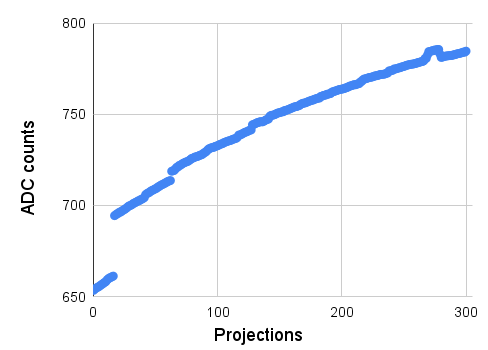} 
         \caption{}
         \label{fig:drift}
     \end{subfigure}
     \caption{A single reconstructed WC sphere of the calibration phantom can be seen in Figure \ref{fig:BallBearing}. The red dashed line through the sphere is where the line out for the central image was taken. Figure \ref{fig:lineout} shows a line out through the sphere. Figure \ref{fig:drift} plots the dark current that was observed on the detector over the time taken to collect a scan data set.}
        \label{fig:XRaydata}
\end{figure}
\\
To understand the image quality and the performance of the imaging system, we first imaged and reconstructed the calibration phantom shown in Figure \ref{fig:Phantom}. The CT image shown in Figure \ref{fig:BallBearing} is a cropped image around a single WC sphere with a red dashed and circular line to show the region-of-interest (ROI). The image was reconstructed using the technique described in the \hyperref[subsec:DataProcessing]{Data Processing and Image Reconstruction} section, to remove some of the noise present in the data. A profile was plotted along the red dashed line and is shown in Figure \ref{fig:lineout}. The line out shows the changing attenuation coefficient across the sphere, with the mean attenuation coefficient within the ROI  measured to be 1.02 x 10$^{-2}$ mm$^{-1}$ and with standard deviation 1.2 x $10^{-3}$ mm$^{-1}$. A cupping artefact is visible in the image of the sphere and its profile,  which is a characteristic of scatter and beam hardening. While the X-ray ICS source presents a narrower X-ray spectrum than a bremsstrahlung source and fewer of the X-ray photons are at a lower energy it is clear that a narrower-still ICS X-ray spectrum is required to remove the beam hardening imaging artifact. We estimated the diameter of the ball bearing as 2.4 mm directly from the CT image. This is consistent with the design parameter of these ball bearings, which were specified to be 2.5 mm. The X-ray images were noisy in part due to a lower photon flux than was suitable for the type of detector being used, and this was compensated by taking multiple X-ray images at each projection. The detector also showed a dark current that drifted during scans (Figure \ref{fig:drift}) due to the detector warming up, although the effect was compensated for by using interleaved $I_d$ frames accompanying each $I$ frame.  
\\
\\
Figure \ref{fig:CTRenSlice} shows a slice taken through the reconstrued  AM part. A line out averaged over several image rows, corresponding to the red box in Figure \ref{fig:CTRenSlice}, was used to measure the full-width-half-maximum thickness of each fin in the profile. The values are approximately 0.45, 0.56, 0.68, 0.68, 1.13, and 1.24 mm. These values were compared against a destructive measurement method, in which the part was cross-sectioned and polished before high-resolution images were taken with a microscope. The destructive method was compared against the XCT method as can be seen in Figure \ref{fig:RenLineout}: the  XCT measurements are on average 15\% larger than the destructive method. This large discrepancy can be accounted for by the fact that in the XCT data it is difficult to determine the size of the outside fins, this is because of edge effects as the AM part was similar in size to our field of view. A fully reconstructed and 3D rendered image of the AM part was produced as shown in Figure \ref{fig:RenReconstruct}. This allowed for easy manipulation and inspection of flaws in the AM part, none were found at the resolution of the current system.

\section*{Discussion}
\label{sec:Discussion} 

We have demonstrated for the first time the imaging and 3D reconstruction of an AM part made from a Nickel 718 alloy, using an LPA-ICS source. These results are only the beginning of this promising new application of a LPA-ICS X-ray source, with the potential to perform high spatial resolution imaging of superalloy parts. This inspection technique has the potential to revolutionise the types of materials and industries that can use  AM parts by ensuring that the standards of the parts meet the required engineering specifications.
\\
\\
While the imaging of the AM part was successful in resolving the object on a macro-scale, the finer detail was not possible to resolve and the initial $<10 \mu m$ resolution, which is needed for AM inspection, was not achieved. While the geometric magnification in theory gave us a $11 \mu m$ pixel size this was not achieved because of the lack of sensitivity of the detector, resulting in a final image resolution of $50 \mu m$. The detector is thus one of the limitations of this experiment: novel detector technology that would be compatible with this new source may be necessary. One potential candidate is the GLO \cite{Cherepy2021} scintillator material lens coupled to a CCD or CMOS camera, which has the potential to offer the correct resolution with the right geometric magnification. More importantly, this detector has a higher X-ray-to-light conversion efficiency than other detectors, allowing us to fully utilise an LPA-ICS X-ray source. 
\\
\\
Future experiments are planned which would focus on achieving higher energy X-rays, of around 1 MeV, which is seen as an area of X-ray energies that is difficult to reach while maintaining a micrometre focal spot size for other X-ray sources. The other focus of future experiments will be on improving X-ray source bandwidth. A bandwidth of 10 \% or less will be required to remove beam hardening in the image - the electron beams required to do this have been demonstrated at other facilities \cite{Maier2020}, indicating that this is achievable. Alternatively, narrowband X-rays can also be generated with the use of a plasma lens \cite{Brummer2022}. Either of these improvements would realise a narrower band, small spot size ICS X-ray source that is required to see the full imaging potential of such an X-ray source.     


\section*{Acknowledgements}
This work was generously funded by the following grants: Engineering Physical Science Research Council (EPSRC) EP/S001379/1 and the U.S. Department of Energy by Lawrence Livermore National Laboratory under Contract DE-AC52-07NA27344. The LLNL team was supported by the LLNL-LDRD Program under Project No. LDRD 20-SI-001. The BELLA LPA-ICS facility is developed with support from US DOE NNSA DNN R \& D (NA-22) funding. User access to the BELLA laser facility was provided by the US DOE FES LaserNetUS user access program, under DOE contract DE-AC02-05CH11231. 

\section*{Author contributions statement}

CT conceived the experiment(s), applied for the beam time and wrote the manuscript with contributions from HM and SK. CT, CA, OF, NT, HM, SG, ND, MS, MF, CG, QC, RJ, KN and JvT conducted the experiment(s), SK and WDB processed the data and reconstructed and analysed the images, QC and R analysed the electron and X-ray spectra. All authors reviewed the manuscript. 
CT conceived the experiment(s), applied for the beam time and wrote the manuscript with contributions from HM and SK. CT, CA, OF, NT, HM, SG, ND, MS, MF, CG, QC, RJ, KN and JvT conducted the experiment(s), SK and WDB processed the data and reconstructed and analysed the images, QC and R analysed the electron and X-ray spectra. All authors reviewed the manuscript. 


\bibliography{sn-bibliography}


\begin{thebibliography}{33}
\ifx \bisbn   \undefined \def \bisbn  #1{ISBN #1}\fi
\ifx \binits  \undefined \def \binits#1{#1}\fi
\ifx \bauthor  \undefined \def \bauthor#1{#1}\fi
\ifx \batitle  \undefined \def \batitle#1{#1}\fi
\ifx \bjtitle  \undefined \def \bjtitle#1{#1}\fi
\ifx \bvolume  \undefined \def \bvolume#1{\textbf{#1}}\fi
\ifx \byear  \undefined \def \byear#1{#1}\fi
\ifx \bissue  \undefined \def \bissue#1{#1}\fi
\ifx \bfpage  \undefined \def \bfpage#1{#1}\fi
\ifx \blpage  \undefined \def \blpage #1{#1}\fi
\ifx \burl  \undefined \def \burl#1{\textsf{#1}}\fi
\ifx \doiurl  \undefined \def \doiurl#1{\url{https://doi.org/#1}}\fi
\ifx \betal  \undefined \def \betal{\textit{et al.}}\fi
\ifx \binstitute  \undefined \def \binstitute#1{#1}\fi
\ifx \binstitutionaled  \undefined \def \binstitutionaled#1{#1}\fi
\ifx \bctitle  \undefined \def \bctitle#1{#1}\fi
\ifx \beditor  \undefined \def \beditor#1{#1}\fi
\ifx \bpublisher  \undefined \def \bpublisher#1{#1}\fi
\ifx \bbtitle  \undefined \def \bbtitle#1{#1}\fi
\ifx \bedition  \undefined \def \bedition#1{#1}\fi
\ifx \bseriesno  \undefined \def \bseriesno#1{#1}\fi
\ifx \blocation  \undefined \def \blocation#1{#1}\fi
\ifx \bsertitle  \undefined \def \bsertitle#1{#1}\fi
\ifx \bsnm \undefined \def \bsnm#1{#1}\fi
\ifx \bsuffix \undefined \def \bsuffix#1{#1}\fi
\ifx \bparticle \undefined \def \bparticle#1{#1}\fi
\ifx \barticle \undefined \def \barticle#1{#1}\fi
\bibcommenthead
\ifx \bconfdate \undefined \def \bconfdate #1{#1}\fi
\ifx \botherref \undefined \def \botherref #1{#1}\fi
\ifx \url \undefined \def \url#1{\textsf{#1}}\fi
\ifx \bchapter \undefined \def \bchapter#1{#1}\fi
\ifx \bbook \undefined \def \bbook#1{#1}\fi
\ifx \bcomment \undefined \def \bcomment#1{#1}\fi
\ifx \oauthor \undefined \def \oauthor#1{#1}\fi
\ifx \citeauthoryear \undefined \def \citeauthoryear#1{#1}\fi
\ifx \endbibitem  \undefined \def \endbibitem {}\fi
\ifx \bconflocation  \undefined \def \bconflocation#1{#1}\fi
\ifx \arxivurl  \undefined \def \arxivurl#1{\textsf{#1}}\fi
\csname PreBibitemsHook\endcsname

\bibitem[\protect\citeauthoryear{Blakey-Milner et~al.}{2021}]{BLAKEYMILNER2021}
\begin{barticle}
\bauthor{\bsnm{Blakey-Milner}, \binits{B.}},
\bauthor{\bsnm{Gradl}, \binits{P.}},
\bauthor{\bsnm{Snedden}, \binits{G.}},
\bauthor{\bsnm{Brooks}, \binits{M.}},
\bauthor{\bsnm{Pitot}, \binits{J.}},
\bauthor{\bsnm{Lopez}, \binits{E.}},
\bauthor{\bsnm{Leary}, \binits{M.}},
\bauthor{\bsnm{Berto}, \binits{F.}},
\bauthor{\bsnm{{du Plessis}}, \binits{A.}}:
\batitle{Metal additive manufacturing in aerospace: A review}.
\bjtitle{Materials \& Design}
\bvolume{209},
\bfpage{110008}
(\byear{2021})
\doiurl{10.1016/j.matdes.2021.110008}
\end{barticle}
\endbibitem

\bibitem[\protect\citeauthoryear{Zhao et~al.}{2023}]{ZHAO2023}
\begin{barticle}
\bauthor{\bsnm{Zhao}, \binits{N.}},
\bauthor{\bsnm{Parthasarathy}, \binits{M.}},
\bauthor{\bsnm{Patil}, \binits{S.}},
\bauthor{\bsnm{Coates}, \binits{D.}},
\bauthor{\bsnm{Myers}, \binits{K.}},
\bauthor{\bsnm{Zhu}, \binits{H.}},
\bauthor{\bsnm{Li}, \binits{W.}}:
\batitle{Direct additive manufacturing of metal parts for automotive applications}.
\bjtitle{Journal of Manufacturing Systems}
\bvolume{68},
\bfpage{368}--\blpage{375}
(\byear{2023})
\doiurl{10.1016/j.jmsy.2023.04.008}
\end{barticle}
\endbibitem

\bibitem[\protect\citeauthoryear{Domenico et~al.}{2023}]{JeongHa2023}
\begin{botherref}
\oauthor{\bsnm{Domenico}, \binits{M.}},
\oauthor{\bsnm{Barbara}, \binits{M.}},
\oauthor{\bsnm{Emanuele}, \binits{V.}},
\oauthor{\bsnm{Stefano}, \binits{F.}},
\oauthor{\bsnm{Federico}, \binits{S.}},
\oauthor{\bsnm{Giovanni}, \binits{T.}},
\oauthor{\bsnm{Marco}, \binits{S.}},
\oauthor{\bsnm{Vito}, \binits{I.}},
\oauthor{\bsnm{Giuseppe}, \binits{M.}},
\oauthor{\bsnm{Jeong-Ha}, \binits{Y.}},
\oauthor{\bsnm{Salvatore}, \binits{G.}},
\oauthor{\bsnm{Antonio}, \binits{L.}},
\oauthor{\bsnm{Massimo}, \binits{M.}},
\oauthor{\bsnm{Ram{\'o}n}, \binits{M.B.}},
\oauthor{\bsnm{Caterina}, \binits{R.}},
\oauthor{\bsnm{Lionel}, \binits{R.}}:
Experiences of additive manufacturing for nuclear fusion applications: The case of the wishbone of the divertor of demo project.
Advances on Mechanics, Design Engineering and Manufacturing IV,
1030--1041
(2023)
\doiurl{10.1007/978-3-031-15928-2_90}
\end{botherref}
\endbibitem

\bibitem[\protect\citeauthoryear{Nordin et~al.}{2017}]{Nordin_2017}
\begin{barticle}
\bauthor{\bsnm{Nordin}, \binits{N.A.B.}},
\bauthor{\bsnm{Johar}, \binits{M.A.B.}},
\bauthor{\bsnm{Ibrahim}, \binits{M.H.I.B.}},
\bauthor{\bsnm{Marwah}, \binits{O.M.F.}}:
\batitle{Advances in high temperature materials for additive manufacturing}.
\bjtitle{IOP Conference Series: Materials Science and Engineering}
\bvolume{226}(\bissue{1}),
\bfpage{012176}
(\byear{2017})
\doiurl{10.1088/1757-899X/226/1/012176}
\end{barticle}
\endbibitem

\bibitem[\protect\citeauthoryear{Niknam et~al.}{2020}]{Niknam2020}
\begin{barticle}
\bauthor{\bsnm{Niknam}, \binits{S.A.}},
\bauthor{\bsnm{Mortazavi}, \binits{M.}},
\bauthor{\bsnm{Li}, \binits{D.}}:
\batitle{Additively manufactured heat exchangers: a review on opportunities and challenges}.
\bjtitle{The International Journal of Advanced Manufacturing Technology}
\bvolume{112},
\bfpage{601}--\blpage{618}
(\byear{2020})
\end{barticle}
\endbibitem

\bibitem[\protect\citeauthoryear{Withers et~al.}{2021}]{Withers2021}
\begin{barticle}
\bauthor{\bsnm{Withers}, \binits{P.J.}},
\bauthor{\bsnm{Bouman}, \binits{C.}},
\bauthor{\bsnm{Carmignato}, \binits{S.}},
\bauthor{\bsnm{Cnudde}, \binits{V.}},
\bauthor{\bsnm{Grimaldi}, \binits{D.}},
\bauthor{\bsnm{Hagen}, \binits{C.K.}},
\bauthor{\bsnm{Maire}, \binits{E.}},
\bauthor{\bsnm{Manley}, \binits{M.}},
\bauthor{\bsnm{Du~Plessis}, \binits{A.}},
\bauthor{\bsnm{Stock}, \binits{S.R.}}:
\batitle{X-ray computed tomography}.
\bjtitle{Nature Reviews Methods Primers}
\bvolume{1}(\bissue{1}),
\bfpage{18}
(\byear{2021})
\doiurl{10.1038/s43586-021-00015-4}
\end{barticle}
\endbibitem

\bibitem[\protect\citeauthoryear{Martz et~al.}{2016}]{Martz2016}
\begin{bbook}
\bauthor{\bsnm{Martz}, \binits{H.E.}},
\bauthor{\bsnm{Logan}, \binits{C.M.}},
\bauthor{\bsnm{Schneberk}, \binits{D.J.}},
\bauthor{\bsnm{Shull}, \binits{P.J.}}:
\bbtitle{X-Ray Imaging: Fundamentals, Industrial Techniques and Applications (1st Ed.)}.
\bpublisher{CRC Press}, \blocation{???}
(\byear{2016})
\end{bbook}
\endbibitem

\bibitem[\protect\citeauthoryear{Thompson et~al.}{2016}]{Thompson_2016}
\begin{barticle}
\bauthor{\bsnm{Thompson}, \binits{A.}},
\bauthor{\bsnm{Maskery}, \binits{I.}},
\bauthor{\bsnm{Leach}, \binits{R.K.}}:
\batitle{X-ray computed tomography for additive manufacturing: a review}.
\bjtitle{Measurement Science and Technology}
\bvolume{27}(\bissue{7}),
\bfpage{072001}
(\byear{2016})
\doiurl{10.1088/0957-0233/27/7/072001}
\end{barticle}
\endbibitem

\bibitem[\protect\citeauthoryear{Azevedo et~al.}{2016}]{Azevedo2016}
\begin{barticle}
\bauthor{\bsnm{Azevedo}, \binits{S.G.}},
\bauthor{\bsnm{Martz}, \binits{H.E.}},
\bauthor{\bsnm{Aufderheide}, \binits{M.B.}},
\bauthor{\bsnm{Brown}, \binits{W.D.}},
\bauthor{\bsnm{Champley}, \binits{K.M.}},
\bauthor{\bsnm{Kallman}, \binits{J.S.}},
\bauthor{\bsnm{Roberson}, \binits{G.P.}},
\bauthor{\bsnm{Schneberk}, \binits{D.}},
\bauthor{\bsnm{Seetho}, \binits{I.M.}},
\bauthor{\bsnm{Smith}, \binits{J.A.}}:
\batitle{System-independent characterization of materials using dual-energy computed tomography}.
\bjtitle{IEEE Transactions on Nuclear Science}
\bvolume{63}(\bissue{1}),
\bfpage{341}--\blpage{350}
(\byear{2016})
\doiurl{10.1109/TNS.2016.2514364}
\end{barticle}
\endbibitem

\bibitem[\protect\citeauthoryear{Martz et~al.}{2017}]{Martz2017}
\begin{barticle}
\bauthor{\bsnm{Martz}, \binits{H.E.}},
\bauthor{\bsnm{Glenn}, \binits{S.M.}},
\bauthor{\bsnm{Smith}, \binits{J.A.}},
\bauthor{\bsnm{Divin}, \binits{C.J.}},
\bauthor{\bsnm{Azevedo}, \binits{S.G.}}:
\batitle{Poly- versus mono-energetic dual-spectrum non-intrusive inspection of cargo containers}.
\bjtitle{IEEE Transactions on Nuclear Science}
\bvolume{64}(\bissue{7}),
\bfpage{1709}--\blpage{1718}
(\byear{2017})
\doiurl{10.1109/TNS.2017.2652455}
\end{barticle}
\endbibitem

\bibitem[\protect\citeauthoryear{Weller et~al.}{2009}]{Weller2009}
\begin{barticle}
\bauthor{\bsnm{Weller}, \binits{H.R.}},
\bauthor{\bsnm{Ahmed}, \binits{M.W.}},
\bauthor{\bsnm{Gao}, \binits{H.}},
\bauthor{\bsnm{Tornow}, \binits{W.}},
\bauthor{\bsnm{Wu}, \binits{Y.K.}},
\bauthor{\bsnm{Gai}, \binits{M.}},
\bauthor{\bsnm{Miskimen}, \binits{R.}}:
\batitle{Research opportunities at the upgraded \uppercase{HI}$\gamma$ \uppercase{S} facility}.
\bjtitle{Progress in Particle and Nuclear Physics}
\bvolume{62}(\bissue{1}),
\bfpage{257}--\blpage{303}
(\byear{2009})
\doiurl{10.1016/j.ppnp.2008.07.001}
\end{barticle}
\endbibitem

\bibitem[\protect\citeauthoryear{Mangles et~al.}{2004}]{Mangles2004}
\begin{barticle}
\bauthor{\bsnm{Mangles}, \binits{S.P.D.}},
\bauthor{\bsnm{Murphy}, \binits{C.D.}},
\bauthor{\bsnm{Najmudin}, \binits{Z.}},
\bauthor{\bsnm{Thomas}, \binits{A.G.R.}},
\bauthor{\bsnm{Collier}, \binits{J.L.}},
\bauthor{\bsnm{Dangor}, \binits{A.E.}},
\bauthor{\bsnm{Divall}, \binits{E.J.}},
\bauthor{\bsnm{Foster}, \binits{P.S.}},
\bauthor{\bsnm{Gallacher}, \binits{J.G.}},
\bauthor{\bsnm{Hooker}, \binits{C.J.}},
\bauthor{\bsnm{Jaroszynski}, \binits{D.A.}},
\bauthor{\bsnm{Langley}, \binits{A.J.}},
\bauthor{\bsnm{Mori}, \binits{W.B.}},
\bauthor{\bsnm{Norreys}, \binits{P.A.}},
\bauthor{\bsnm{Tsung}, \binits{F.S.}},
\bauthor{\bsnm{Viskup}, \binits{R.}},
\bauthor{\bsnm{Walton}, \binits{B.R.}},
\bauthor{\bsnm{Krushelnick}, \binits{K.}}:
\batitle{Monoenergetic beams of relativistic electrons from intense laser--plasma interactions}.
\bjtitle{Nature}
\bvolume{431}(\bissue{7008}),
\bfpage{535}--\blpage{538}
(\byear{2004})
\doiurl{10.1038/nature02939}
\end{barticle}
\endbibitem

\bibitem[\protect\citeauthoryear{Geddes et~al.}{2004}]{Geddes2004}
\begin{barticle}
\bauthor{\bsnm{Geddes}, \binits{C.G.R.}},
\bauthor{\bsnm{Toth}, \binits{C.}},
\bauthor{\bsnm{Tilborg}, \binits{J.}},
\bauthor{\bsnm{Esarey}, \binits{E.}},
\bauthor{\bsnm{Schroeder}, \binits{C.B.}},
\bauthor{\bsnm{Bruhwiler}, \binits{D.}},
\bauthor{\bsnm{Nieter}, \binits{C.}},
\bauthor{\bsnm{Cary}, \binits{J.}},
\bauthor{\bsnm{Leemans}, \binits{W.P.}}:
\batitle{High-quality electron beams from a laser wakefield accelerator using plasma-channel guiding}.
\bjtitle{Nature}
\bvolume{431}(\bissue{7008}),
\bfpage{538}--\blpage{541}
(\byear{2004})
\doiurl{10.1038/nature02900}
\end{barticle}
\endbibitem

\bibitem[\protect\citeauthoryear{Faure et~al.}{2004}]{Faure2004}
\begin{barticle}
\bauthor{\bsnm{Faure}, \binits{J.}},
\bauthor{\bsnm{Glinec}, \binits{Y.}},
\bauthor{\bsnm{Pukhov}, \binits{A.}},
\bauthor{\bsnm{Kiselev}, \binits{S.}},
\bauthor{\bsnm{Gordienko}, \binits{S.}},
\bauthor{\bsnm{Lefebvre}, \binits{E.}},
\bauthor{\bsnm{Rousseau}, \binits{J.-P.}},
\bauthor{\bsnm{Burgy}, \binits{F.}},
\bauthor{\bsnm{Malka}, \binits{V.}}:
\batitle{A laser--plasma accelerator producing monoenergetic electron beams}.
\bjtitle{Nature}
\bvolume{431}(\bissue{7008}),
\bfpage{541}--\blpage{544}
(\byear{2004})
\doiurl{10.1038/nature02963}
\end{barticle}
\endbibitem

\bibitem[\protect\citeauthoryear{Bingham and Trines}{2014}]{Bingham2014}
\begin{barticle}
\bauthor{\bsnm{Bingham}, \binits{R.}},
\bauthor{\bsnm{Trines}, \binits{E.}}:
\batitle{Introduction to plasma accelerators: the basics}.
\bjtitle{In Proceedings of the CAS-CERN Accelerator School: Plasma Wake Acceleration}
(\byear{2014})
\doiurl{10.5170/CERN-2016-001.67 2}
\end{barticle}
\endbibitem

\bibitem[\protect\citeauthoryear{Plateau et~al.}{2012}]{Plateau2012}
\begin{barticle}
\bauthor{\bsnm{Plateau}, \binits{G.R.}},
\bauthor{\bsnm{Geddes}, \binits{C.G.R.}},
\bauthor{\bsnm{Thorn}, \binits{D.B.}},
\bauthor{\bsnm{Chen}, \binits{M.}},
\bauthor{\bsnm{Benedetti}, \binits{C.}},
\bauthor{\bsnm{Esarey}, \binits{E.}},
\bauthor{\bsnm{Gonsalves}, \binits{A.J.}},
\bauthor{\bsnm{Matlis}, \binits{N.H.}},
\bauthor{\bsnm{Nakamura}, \binits{K.}},
\bauthor{\bsnm{Schroeder}, \binits{C.B.}},
\bauthor{\bsnm{Shiraishi}, \binits{S.}},
\bauthor{\bsnm{Sokollik}, \binits{T.}},
\bauthor{\bsnm{Tilborg}, \binits{J.}},
\bauthor{\bsnm{Toth}, \binits{C.}},
\bauthor{\bsnm{Trotsenko}, \binits{S.}},
\bauthor{\bsnm{Kim}, \binits{T.S.}},
\bauthor{\bsnm{Battaglia}, \binits{M.}},
\bauthor{\bsnm{St\"ohlker}, \binits{T.}},
\bauthor{\bsnm{Leemans}, \binits{W.P.}}:
\batitle{Low-emittance electron bunches from a laser-plasma accelerator measured using single-shot x-ray spectroscopy}.
\bjtitle{Phys. Rev. Lett.}
\bvolume{109},
\bfpage{064802}
(\byear{2012})
\doiurl{10.1103/PhysRevLett.109.064802}
\end{barticle}
\endbibitem

\bibitem[\protect\citeauthoryear{Tsai et~al.}{2015}]{HaiEn2015}
\begin{barticle}
\bauthor{\bsnm{Tsai}, \binits{H.-E.}},
\bauthor{\bsnm{Wang}, \binits{X.}},
\bauthor{\bsnm{Shaw}, \binits{J.M.}},
\bauthor{\bsnm{Li}, \binits{Z.}},
\bauthor{\bsnm{Arefiev}, \binits{A.V.}},
\bauthor{\bsnm{Zhang}, \binits{X.}},
\bauthor{\bsnm{Zgadzaj}, \binits{R.}},
\bauthor{\bsnm{Henderson}, \binits{W.}},
\bauthor{\bsnm{Khudik}, \binits{V.}},
\bauthor{\bsnm{Shvets}, \binits{G.}},
\bauthor{\bsnm{Downer}, \binits{M.C.}}:
\batitle{{Compact tunable Compton x-ray source from laser-plasma accelerator and plasma mirror}}.
\bjtitle{Physics of Plasmas}
\bvolume{22}(\bissue{2}),
\bfpage{023106}
(\byear{2015})
\doiurl{10.1063/1.4907655}
{\href{https://arxiv.org/abs/https://pubs.aip.org/aip/pop/article-pdf/doi/10.1063/1.4907655/16143818/023106\_1\_online.pdf}{{https://pubs.aip.org/aip/pop/article-pdf/doi/10.1063/1.4907655/16143818/023106\_1\_online.pdf}}}
\end{barticle}
\endbibitem

\bibitem[\protect\citeauthoryear{Ta~Phuoc et~al.}{2012}]{Ta_Phuoc2012}
\begin{barticle}
\bauthor{\bsnm{Ta~Phuoc}, \binits{K.}},
\bauthor{\bsnm{Corde}, \binits{S.}},
\bauthor{\bsnm{Thaury}, \binits{C.}},
\bauthor{\bsnm{Malka}, \binits{V.}},
\bauthor{\bsnm{Tafzi}, \binits{A.}},
\bauthor{\bsnm{Goddet}, \binits{J.P.}},
\bauthor{\bsnm{Shah}, \binits{R.C.}},
\bauthor{\bsnm{Sebban}, \binits{S.}},
\bauthor{\bsnm{Rousse}, \binits{A.}}:
\batitle{All-optical compton gamma-ray source}.
\bjtitle{Nature Photonics}
\bvolume{6}(\bissue{5}),
\bfpage{308}--\blpage{311}
(\byear{2012})
\doiurl{10.1038/nphoton.2012.82}
\end{barticle}
\endbibitem

\bibitem[\protect\citeauthoryear{Geddes et~al.}{2015}]{GEDDES2015116}
\begin{barticle}
\bauthor{\bsnm{Geddes}, \binits{C.G.R.}},
\bauthor{\bsnm{Rykovanov}, \binits{S.}},
\bauthor{\bsnm{Matlis}, \binits{N.H.}},
\bauthor{\bsnm{Steinke}, \binits{S.}},
\bauthor{\bsnm{Vay}, \binits{J.-L.}},
\bauthor{\bsnm{Esarey}, \binits{E.H.}},
\bauthor{\bsnm{Ludewigt}, \binits{B.}},
\bauthor{\bsnm{Nakamura}, \binits{K.}},
\bauthor{\bsnm{Quiter}, \binits{B.J.}},
\bauthor{\bsnm{Schroeder}, \binits{C.B.}},
\bauthor{\bsnm{Toth}, \binits{C.}},
\bauthor{\bsnm{Leemans}, \binits{W.P.}}:
\batitle{Compact quasi-monoenergetic photon sources from laser-plasma accelerators for nuclear detection and characterization}.
\bjtitle{Nuclear Instruments and Methods in Physics Research Section B: Beam Interactions with Materials and Atoms}
\bvolume{350},
\bfpage{116}--\blpage{121}
(\byear{2015})
\doiurl{10.1016/j.nimb.2015.01.013}
\end{barticle}
\endbibitem

\bibitem[\protect\citeauthoryear{Kr{\"a}mer et~al.}{2018}]{Kramer2018}
\begin{barticle}
\bauthor{\bsnm{Kr{\"a}mer}, \binits{J.M.}},
\bauthor{\bsnm{Jochmann}, \binits{A.}},
\bauthor{\bsnm{Budde}, \binits{M.}},
\bauthor{\bsnm{Bussmann}, \binits{M.}},
\bauthor{\bsnm{Couperus}, \binits{J.P.}},
\bauthor{\bsnm{Cowan}, \binits{T.E.}},
\bauthor{\bsnm{Debus}, \binits{A.}},
\bauthor{\bsnm{K{\"o}hler}, \binits{A.}},
\bauthor{\bsnm{Kuntzsch}, \binits{M.}},
\bauthor{\bsnm{Laso~Garc{\'\i}a}, \binits{A.}},
\bauthor{\bsnm{Lehnert}, \binits{U.}},
\bauthor{\bsnm{Michel}, \binits{P.}},
\bauthor{\bsnm{Pausch}, \binits{R.}},
\bauthor{\bsnm{Zarini}, \binits{O.}},
\bauthor{\bsnm{Schramm}, \binits{U.}},
\bauthor{\bsnm{Irman}, \binits{A.}}:
\batitle{Making spectral shape measurements in inverse compton scattering a tool for advanced diagnostic applications}.
\bjtitle{Scientific Reports}
\bvolume{8}(\bissue{1}),
\bfpage{1398}
(\byear{2018})
\doiurl{10.1038/s41598-018-19546-0}
\end{barticle}
\endbibitem

\bibitem[\protect\citeauthoryear{Rykovanov et~al.}{2014}]{Rykovanov_2014}
\begin{barticle}
\bauthor{\bsnm{Rykovanov}, \binits{S.G.}},
\bauthor{\bsnm{Geddes}, \binits{C.G.R.}},
\bauthor{\bsnm{Vay}, \binits{J.-L.}},
\bauthor{\bsnm{Schroeder}, \binits{C.B.}},
\bauthor{\bsnm{Esarey}, \binits{E.}},
\bauthor{\bsnm{Leemans}, \binits{W.P.}}:
\batitle{Quasi-monoenergetic femtosecond photon sources from thomson scattering using laser plasma accelerators and plasma channels}.
\bjtitle{Journal of Physics B: Atomic, Molecular and Optical Physics}
\bvolume{47}(\bissue{23}),
\bfpage{234013}
(\byear{2014})
\doiurl{10.1088/0953-4075/47/23/234013}
\end{barticle}
\endbibitem

\bibitem[\protect\citeauthoryear{Döpp et~al.}{2016}]{Döpp_2016}
\begin{barticle}
\bauthor{\bsnm{Döpp}, \binits{A.}},
\bauthor{\bsnm{Guillaume}, \binits{E.}},
\bauthor{\bsnm{Thaury}, \binits{C.}},
\bauthor{\bsnm{Gautier}, \binits{J.}},
\bauthor{\bsnm{Andriyash}, \binits{I.}},
\bauthor{\bsnm{Lifschitz}, \binits{A.}},
\bauthor{\bsnm{Malka}, \binits{V.}},
\bauthor{\bsnm{Rousse}, \binits{A.}},
\bauthor{\bsnm{Phuoc}, \binits{K.T.}}:
\batitle{An all-optical compton source for single-exposure x-ray imaging}.
\bjtitle{Plasma Physics and Controlled Fusion}
\bvolume{58}(\bissue{3}),
\bfpage{034005}
(\byear{2016})
\doiurl{10.1088/0741-3335/58/3/034005}
\end{barticle}
\endbibitem

\bibitem[\protect\citeauthoryear{Haden et~al.}{2016}]{Haden2016}
\begin{bchapter}
\bauthor{\bsnm{Haden}, \binits{D.}},
\bauthor{\bsnm{Chen}, \binits{S.}},
\bauthor{\bsnm{Zhao}, \binits{B.}},
\bauthor{\bsnm{Zhang}, \binits{P.}},
\bauthor{\bsnm{Golovin}, \binits{G.}},
\bauthor{\bsnm{Yan}, \binits{W.}},
\bauthor{\bsnm{Fruhling}, \binits{C.}},
\bauthor{\bsnm{Banerjee}, \binits{S.}},
\bauthor{\bsnm{Umstadter}, \binits{D.}}:
\bctitle{{High-resolution radiography of thick steel objects using an all-laser-driven MeV-energy x-ray source}}.
In: \beditor{\bsnm{Khounsary}, \binits{A.M.}},
\beditor{\bsnm{Dorssen}, \binits{G.E.}} (eds.)
\bbtitle{Advances in Laboratory-based X-Ray Sources, Optics, and Applications V},
vol. \bseriesno{9964},
p. \bfpage{99640}.
\bpublisher{SPIE}, \blocation{???}
(\byear{2016}).
\doiurl{10.1117/12.2241606} .
\bcomment{International Society for Optics and Photonics}.
\burl{https://doi.org/10.1117/12.2241606}
\end{bchapter}
\endbibitem

\bibitem[\protect\citeauthoryear{Chen et~al.}{2016}]{CHEN2016}
\begin{barticle}
\bauthor{\bsnm{Chen}, \binits{S.}},
\bauthor{\bsnm{Golovin}, \binits{G.}},
\bauthor{\bsnm{Miller}, \binits{C.}},
\bauthor{\bsnm{Haden}, \binits{D.}},
\bauthor{\bsnm{Banerjee}, \binits{S.}},
\bauthor{\bsnm{Zhang}, \binits{P.}},
\bauthor{\bsnm{Liu}, \binits{C.}},
\bauthor{\bsnm{Zhang}, \binits{J.}},
\bauthor{\bsnm{Zhao}, \binits{B.}},
\bauthor{\bsnm{Clarke}, \binits{S.}},
\bauthor{\bsnm{Pozzi}, \binits{S.}},
\bauthor{\bsnm{Umstadter}, \binits{D.}}:
\batitle{Shielded radiography with a laser-driven mev-energy x-ray source}.
\bjtitle{Nuclear Instruments and Methods in Physics Research Section B: Beam Interactions with Materials and Atoms}
\bvolume{366},
\bfpage{217}--\blpage{223}
(\byear{2016})
\doiurl{10.1016/j.nimb.2015.11.007}
\end{barticle}
\endbibitem

\bibitem[\protect\citeauthoryear{Ma et~al.}{2020}]{Ma2020}
\begin{barticle}
\bauthor{\bsnm{Ma}, \binits{Y.}},
\bauthor{\bsnm{Hua}, \binits{J.}},
\bauthor{\bsnm{Liu}, \binits{D.}},
\bauthor{\bsnm{He}, \binits{Y.}},
\bauthor{\bsnm{Zhang}, \binits{T.}},
\bauthor{\bsnm{Chen}, \binits{J.}},
\bauthor{\bsnm{Yang}, \binits{F.}},
\bauthor{\bsnm{Ning}, \binits{X.}},
\bauthor{\bsnm{Yang}, \binits{Z.}},
\bauthor{\bsnm{Zhang}, \binits{J.}},
\bauthor{\bsnm{Pai}, \binits{C.-H.}},
\bauthor{\bsnm{Gu}, \binits{Y.}},
\bauthor{\bsnm{Lu}, \binits{W.}}:
\batitle{Region-of-interest micro-focus computed tomography based on an all-optical inverse compton scattering source}.
\bjtitle{Matter and Radiation at Extremes}
\bvolume{5}(\bissue{6}),
\bfpage{064401}
(\byear{2020})
\doiurl{10.1063/5.0016034}
{\href{https://arxiv.org/abs/https://pubs.aip.org/aip/mre/article-pdf/doi/10.1063/5.0016034/15785994/064401\_1\_online.pdf}{{https://pubs.aip.org/aip/mre/article-pdf/doi/10.1063/5.0016034/15785994/064401\_1\_online.pdf}}}
\end{barticle}
\endbibitem

\bibitem[\protect\citeauthoryear{{Ostermayr} et~al.}{2020}]{Ostermayr2020}
\begin{bchapter}
\bauthor{\bsnm{{Ostermayr}}, \binits{T.}},
\bauthor{\bsnm{{Tsai}}, \binits{H.-E.}},
\bauthor{\bsnm{{Ettelbrick}}, \binits{R.}},
\bauthor{\bsnm{{Fan-Chiang}}, \binits{L.}},
\bauthor{\bsnm{{Jacob}}, \binits{R.}},
\bauthor{\bsnm{{Laut}}, \binits{A.}},
\bauthor{\bsnm{{Zhou}}, \binits{O.}},
\bauthor{\bsnm{{van Tilborg}}, \binits{J.}},
\bauthor{\bsnm{{Isono}}, \binits{F.}},
\bauthor{\bsnm{{Barber}}, \binits{S.K.}},
\bauthor{\bsnm{{Lehe}}, \binits{R.}},
\bauthor{\bsnm{{Vay}}, \binits{J.-L.}},
\bauthor{\bsnm{{Gonsalves}}, \binits{A.}},
\bauthor{\bsnm{{Nakamura}}, \binits{K.}},
\bauthor{\bsnm{{Toth}}, \binits{C.}},
\bauthor{\bsnm{{Schroeder}}, \binits{C.}},
\bauthor{\bsnm{{Geddes}}, \binits{C.}},
\bauthor{\bsnm{{Esarey}}, \binits{E.}}:
\bctitle{The bella center hundred terawatt laser system for photon sources and user experiments}.
In: \bbtitle{APS Division of Plasma Physics Meeting Abstracts}.
\bsertitle{APS Meeting Abstracts},
vol. \bseriesno{2020},
pp. \bfpage{04}--\blpage{002}
(\byear{2020})
\end{bchapter}
\endbibitem

\bibitem[\protect\citeauthoryear{Chen et~al.}{2023}]{Chen2023}
\begin{bchapter}
\bauthor{\bsnm{Chen}, \binits{Q.}},
\bauthor{\bsnm{Jacob}, \binits{R.}},
\bauthor{\bsnm{Tilborg}, \binits{J.}},
\bauthor{\bsnm{Gonsalves}, \binits{A.}},
\bauthor{\bsnm{Nakamura}, \binits{K.}},
\bauthor{\bsnm{Schroeder}, \binits{C.}},
\bauthor{\bsnm{Esarey}, \binits{E.}},
\bauthor{\bsnm{Geddes}, \binits{C.}}:
\bctitle{Development of the mev thomson-scattered gamma ray source using laser plasma accelerators at the bella center}.
In: \beditor{\bsnm{Menoni}, \binits{C.S.}},
\beditor{\bsnm{Nejdl}, \binits{J.}} (eds.)
\bbtitle{Compact Radiation Sources from EUV to Gamma-rays: Development and Applications},
vol. \bseriesno{PC12582},
p. \bfpage{125820}.
\bpublisher{SPIE}, \blocation{???}
(\byear{2023}).
\doiurl{10.1117/12.2665755} .
\bcomment{International Society for Optics and Photonics}.
\burl{https://doi.org/10.1117/12.2665755}
\end{bchapter}
\endbibitem

\bibitem[\protect\citeauthoryear{Zhou et~al.}{2021}]{Ocean2021}
\begin{barticle}
\bauthor{\bsnm{Zhou}, \binits{O.}},
\bauthor{\bsnm{Tsai}, \binits{H.-E.}},
\bauthor{\bsnm{Ostermayr}, \binits{T.M.}},
\bauthor{\bsnm{Fan-Chiang}, \binits{L.}},
\bauthor{\bsnm{Tilborg}, \binits{J.}},
\bauthor{\bsnm{Schroeder}, \binits{C.B.}},
\bauthor{\bsnm{Esarey}, \binits{E.}},
\bauthor{\bsnm{Geddes}, \binits{C.G.R.}}:
\batitle{{Effect of nozzle curvature on supersonic gas jets used in laser–plasma acceleration}}.
\bjtitle{Physics of Plasmas}
\bvolume{28}(\bissue{9}),
\bfpage{093107}
(\byear{2021})
\doiurl{10.1063/5.0058963}
{\href{https://arxiv.org/abs/https://pubs.aip.org/aip/pop/article-pdf/doi/10.1063/5.0058963/13421446/093107\_1\_online.pdf}{{https://pubs.aip.org/aip/pop/article-pdf/doi/10.1063/5.0058963/13421446/093107\_1\_online.pdf}}}
\end{barticle}
\endbibitem

\bibitem[\protect\citeauthoryear{Swinehart}{1962}]{BeerLambert}
\begin{barticle}
\bauthor{\bsnm{Swinehart}, \binits{D.F.}}:
\batitle{The beer-lambert law}.
\bjtitle{Journal of Chemical Education}
\bvolume{39}(\bissue{7}),
\bfpage{333}
(\byear{1962})
\doiurl{10.1021/ed039p333}
\end{barticle}
\endbibitem

\bibitem[\protect\citeauthoryear{Champley et~al.}{2022}]{Kyle2022}
\begin{barticle}
\bauthor{\bsnm{Champley}, \binits{K.M.}},
\bauthor{\bsnm{Willey}, \binits{T.M.}},
\bauthor{\bsnm{Kim}, \binits{H.}},
\bauthor{\bsnm{Bond}, \binits{K.}},
\bauthor{\bsnm{Glenn}, \binits{S.M.}},
\bauthor{\bsnm{Smith}, \binits{J.A.}},
\bauthor{\bsnm{Kallman}, \binits{J.S.}},
\bauthor{\bsnm{Brown}, \binits{W.D.}},
\bauthor{\bsnm{Seetho}, \binits{I.M.}},
\bauthor{\bsnm{Keene}, \binits{L.}},
\bauthor{\bsnm{Azevedo}, \binits{S.G.}},
\bauthor{\bsnm{McMichael}, \binits{L.D.}},
\bauthor{\bsnm{Overturf}, \binits{G.}},
\bauthor{\bsnm{Martz}, \binits{H.E.}}:
\batitle{Livermore tomography tools: Accurate, fast, and flexible software for tomographic science}.
\bjtitle{NDT \& E International}
\bvolume{126},
\bfpage{102595}
(\byear{2022})
\doiurl{10.1016/j.ndteint.2021.102595}
\end{barticle}
\endbibitem

\bibitem[\protect\citeauthoryear{Cherepy et~al.}{2021}]{Cherepy2021}
\begin{barticle}
\bauthor{\bsnm{Cherepy}, \binits{N.J.}},
\bauthor{\bsnm{Seeley}, \binits{Z.M.}},
\bauthor{\bsnm{Schneberk}, \binits{D.J.}},
\bauthor{\bsnm{McNamee}, \binits{C.J.}},
\bauthor{\bsnm{Pincus}, \binits{C.I.}},
\bauthor{\bsnm{Rudzik}, \binits{T.J.}},
\bauthor{\bsnm{Osborne}, \binits{R.A.}},
\bauthor{\bsnm{Phillips}, \binits{I.R.}},
\bauthor{\bsnm{Nicolino}, \binits{J.A.}},
\bauthor{\bsnm{Hall}, \binits{J.}},
\bauthor{\bsnm{Townsend}, \binits{A.}},
\bauthor{\bsnm{Payne}, \binits{S.A.}}:
\batitle{{Lens-coupled MeV x-radiography and CT with transparent ceramic GLO scintillators}}.
\bjtitle{Hard X-Ray, Gamma-Ray, and Neutron Detector Physics XXIII}
\bvolume{11838},
\bfpage{118380}
(\byear{2021})
\doiurl{10.1117/12.2595956} .
\bcomment{International Society for Optics and Photonics}
\end{barticle}
\endbibitem

\bibitem[\protect\citeauthoryear{Maier et~al.}{2020}]{Maier2020}
\begin{barticle}
\bauthor{\bsnm{Maier}, \binits{A.R.}},
\bauthor{\bsnm{Delbos}, \binits{N.M.}},
\bauthor{\bsnm{Eichner}, \binits{T.}},
\bauthor{\bsnm{H\"ubner}, \binits{L.}},
\bauthor{\bsnm{Jalas}, \binits{S.}},
\bauthor{\bsnm{Jeppe}, \binits{L.}},
\bauthor{\bsnm{Jolly}, \binits{S.W.}},
\bauthor{\bsnm{Kirchen}, \binits{M.}},
\bauthor{\bsnm{Leroux}, \binits{V.}},
\bauthor{\bsnm{Messner}, \binits{P.}},
\bauthor{\bsnm{Schnepp}, \binits{M.}},
\bauthor{\bsnm{Trunk}, \binits{M.}},
\bauthor{\bsnm{Walker}, \binits{P.A.}},
\bauthor{\bsnm{Werle}, \binits{C.}},
\bauthor{\bsnm{Winkler}, \binits{P.}}:
\batitle{Decoding sources of energy variability in a laser-plasma accelerator}.
\bjtitle{Phys. Rev. X}
\bvolume{10},
\bfpage{031039}
(\byear{2020})
\doiurl{10.1103/PhysRevX.10.031039}
\end{barticle}
\endbibitem

\bibitem[\protect\citeauthoryear{Br{\"u}mmer et~al.}{2022}]{Brummer2022}
\begin{barticle}
\bauthor{\bsnm{Br{\"u}mmer}, \binits{T.}},
\bauthor{\bsnm{Bohlen}, \binits{S.}},
\bauthor{\bsnm{Gr{\"u}ner}, \binits{F.}},
\bauthor{\bsnm{Osterhoff}, \binits{J.}},
\bauthor{\bsnm{P{\~o}der}, \binits{K.}}:
\batitle{Compact all-optical precision-tunable narrowband hard compton x-ray source}.
\bjtitle{Scientific Reports}
\bvolume{12}(\bissue{1}),
\bfpage{16017}
(\byear{2022})
\doiurl{10.1038/s41598-022-20283-8}
\end{barticle}
\endbibitem

\end{thebibliography}

\end{document}